\documentclass[12pt, a4paper]{article}
\usepackage{latexsym}
\usepackage{amsmath}
\usepackage{amssymb}
\usepackage{mathrsfs}
\usepackage{amsthm}
\usepackage{amsfonts}
\usepackage{caption}
\usepackage{xcolor} 
\usepackage{color}
\usepackage{rotating}
\usepackage{graphicx}
\usepackage{subfig}
\usepackage{dsfont}

\usepackage[T1]{fontenc}
\usepackage[round]{natbib}

\def \dsE {\text{$\mathds{E}$}}

 \DeclareMathOperator{\Var}{Var}

 \DeclareMathOperator{\RMSE}{RMSE}

\def \fvec {\text{\boldmath$f$}}

\def \xvec {\text{\boldmath$x$}}    
\def \yvec {\text{\boldmath$y$}}    
\def \zvec {\text{\boldmath$z$}}

\def \varepsilonvec   {\text{\boldmath$\varepsilon$}}

\begin{document}

\title{Bayesian Geoadditive Expectile Regression}


\author{Elisabeth Waldmann, Fabian Sobotka, Thomas Kneib}
\maketitle

\begin{abstract}
Regression classes modeling more than the mean of the response have found a lot of attention in the last years. Expectile regression is a special and computationally convenient case of this family of models. Expectiles offer a quantile-like characterisation of a complete distribution and include the mean as a special case. In the frequentist framework the impact of a lot of covariates with very different structures have been made possible. We propose Bayesian expectile regression based on the asymmetric normal distribution. This renders possible incorporating for example linear, nonlinear, spatial and random effects in one model. Furthermore a detailed inference on the estimated parameters can be conducted. Proposal densities based on iterativly weighted least squares updates for the resulting Markov chain Monte Carlo (MCMC) simulation algorithm are proposed and the potential of the approach for extending the flexibility of expectile regression towards complex semiparametric regression specifications is discussed.
\end{abstract}

{\it Keywords: Expectile Regression, Bayesian Semiparametric Regression, Markov random fields, $P$-splines, asymmetric normal distribution, Markov chain Monte Carlo Simulation}


\section{Introduction}
Recent interest in the development of flexible regression specifications has had a specific focus on describing more complex features of the response distribution than only the mean. The standard instrument in this situation is quantile regression \citep{KoeBas1978} where conditional quantiles are related to a regression predictor. A lot of work has been done to extend the simple linear quantile regression model to more advanced approaches like quantile smoothing splines \citep{KoeNgPor1994}, quantile regression for clustered data \citep{ReiBonWan2010} or geoadditive models \citep{Fenske_Kneib_Hothorn_2011}.

Computationally regression quantiles are obtained by minimising an asymmetrically weighted absolute residuals criterion
\begin{equation}\label{eq:quantreg}
 \sum_{i=1}^n w_{\tau}(y_i,\eta_{i\tau})|y_i-\eta_{i\tau}|
\end{equation}
with asymmetric weights
\[
 w_{\tau}(y_i, \eta_{i\tau}) = \begin{cases}
 1-\tau & y_i \le \eta_{i\tau}\\
 \tau & y_i > \eta_{i\tau},
 \end{cases}
\]
a response $y$ and a quantile-specific predictor $\eta_\tau$. This loss function induces additional complexity compared to standard least squares optimisation. As a consequence, expectile regression \citep{NewPow87} that relies on asymmetrically weighted squared residuals
\begin{equation}\label{kneib:eq:expectiles}
 \sum_{i=1}^n w_{\tau}(y_i,\eta_{i\tau})(y_i-\eta_{i\tau})^2
\end{equation} 
has gained considerable interest since expectile regression estimates can be obtained by simple iteratively weighted least squares fits. Extensions to more complicated models have been explored in recent publications for the smoothing of a nonlinear effect \citep{Schnabel:2009}, for geoadditive models \citep{Sobotka:2010} and for instrumental variables \citep{SobotkaMarra:2013}. While basic asymptotic results are available for a least squares estimate \citep[see][]{SobotkaKau:2012}, alternative estimation methods like boosting as introduced to expectiles by \cite{Sobotka:2010} rely on a bootstrap for further inference. An autoregressive definition of expectiles was even introduced for time series analysis \citep{taylor:2008}. In this paper, we introduce a Bayesian formulation of expectile regression that relies on the asymmetric normal distribution (AND) as auxiliary response distribution. The approach is very similar to the estimation of Bayesian quantile regression, where an asymmetric Laplace distribution (ALD) is used instead of the AND.  For detailed information see \cite{Yue}, \cite{kozumi_kobayashi} or \cite{Reed_Yu}. In the case of the AND proposal densities based on iteratively weighted least squares updates for the resulting Markov chain Monte Carlo (MCMC) simulation algorithm are needed.

As an illustrative example, we present a data set dealing with malnutrition in Tanzania. The dependent variable is the so called {\it z-score} of {\it stunting} (a score measuring the height of the child in comparison to a reference population). The latter is the dependent variable and shall be explained by continous covariates like {\it maternal BMI at birth}, {\it age of the child} and categorical covariates ({\it mother's work}, {\it mother's education} and {\it mother's residence}, denoted by $\boldsymbol{X}$). The impact of the continuous covariates used for the explanation of the dependent variable {\it z-score} is not linear thus we use splines. As Tanzania consists of 20 regions over which economic and political situation differ we will also to incorporate the regions into the model. Therefore use a geoadditive model of the type

\begin{equation}
\textnormal{stunting}_i = f(\textnormal{BMI}_i) + f(\textnormal{age}_i) + f_{geo}(\textnormal{region}_i) + \boldsymbol{x_i}\boldsymbol{\beta} + \varepsilon_i,
\label{mod1}
\end{equation}

where the $f$ denotes the nonlinear effects, $\boldsymbol{X}$ contains categorical covariates and $f_{geo}$ is the spatial effect of the different regions of Tanzania.
The necessity of using a model different to mean regression becomes obvious when taking a look at the data: the conditional distribution of the z-score is neither homoscedastic nor symmetric.

The rest of the paper is structured as follows: in the second section we describe the basic ideas of expectile regression and give an overview over the concept of semiparametric regression. We then introduce the above mentioned asymmetric normal distribution and describe the Bayesian algorithm in more detail. The third section contains simulations which study point estimation as well as confidence intervals for the parameters. In Section~\ref{sec:ex} we will describe the above mentioned data set on childhood malnutrition in Tanzania and explain the impact of the different covariates. In the last section we conclude and give an outlook on future plans.
\section{Bayesian Expectile Regression}\label{sec:bayes}
\subsection{Expectile Regression}
Suppose that regression data ($y_i,\boldsymbol{z}_i)$, $i=1,\ldots,n,$ on a continuous response variable $y$ and a covariate vector $\boldsymbol{z}$ are given and shall be analysed in a regression model of the form
\[
 y_i = \eta_{i\tau}+\varepsilon_{i\tau}
\]
where $\eta_{\tau}$ is a predictor formed by the covariates and $\varepsilon_{\tau}$ is an appropriate error term. Unlike in mean regression where regression effects on the mean are of interest, we focus on situations where specific outer parts of the response distribution shall be studied. We will denote the extremeness of these outer parts by the asymmetry parameter $\tau\in(0,1)$ where $\tau=0.5$ corresponds to the central part of the distribution while $\tau\rightarrow0$ and $\tau\rightarrow1$ yield the lower and upper part of the distribution, respectively. The standard approach for implementing such regression models is quantile regression where we assume that the $\tau$-quantile of the error distribution equals zero, i.e.
\[
 P(\varepsilon_{i\tau}\le0)=\tau.
\]
This implies that the predictor $\eta_{i\tau}$ corresponds to the $\tau$-quantile of the response $y_i$ and the regression model can be estimated by minimising the loss function~\eqref{eq:quantreg}.
As an alternative, we will instead focus on the criterion~\eqref{kneib:eq:expectiles} that yields expectile regression estimates. This criterion has the advantage to be differentiable with respect to the regression predictor so that estimates can be obtained by iteratively weighted least squares estimation. Basically, expectiles are an alternative possibility to characterise the distribution of a continuous random variable where $\tau$ indicates the ``extremeness'' of the part of the distribution that shall be studied, see \cite{NewPow87}.

A usual objection against expectiles as compared to quantiles is their lack of an immediate interpretation. While for quantiles the property that $\tau \cdot 100$ percent of the data lie below the regression line and $(1-\tau)$100 percent of the data lie above the regression line is easy to understand, the extremeness of expectiles is hard to transfer to such an easy statement. However, interpretation of expectiles is still possible in the following ways:
\begin{itemize}
\item For i.i.d. data $y_1,\ldots,y_n$, the resulting expectile estimate $\hat{e}_\tau$ will be a weighted average
\[
 \hat{e}_\tau=\sum_{i=1}^nw_iy_i
\]
where the weights $w_i$ depend on the estimated expectile. As a consequence, regression expectiles can also be considered such a weighted average conditioned on a specific covariate vector.
\item Expectiles are tail expectations, i.e. the $\tau$-expectile fulfills
\[
 \tau = \frac{\int_{-\infty}^{e_\tau} |y-e_\tau|f(y)dy}{\int_{-\infty}^\infty |y-e_\tau|f(y)dy}
\]
showing that $e_\tau$ is characterised by a partial moment condition.
\item Usually, one would not only estimate one single expectile but a whole set of expectiles for various values of $\tau$. The collection of all estimates then gives an intuitive impression about the shape of the conditional distribution of the response and in particular allows to detect features such as heteroscedasticity, skewness or kurtosis. Moreover, conditional quantiles can still be calculated from a set of expectiles if quantile estimates are of ultimate interest, as shown by \cite{Efron:91} and refined in \cite{SchuWalSob:2013}.
\item Expectiles are increasingly important when it comes to measuring risks. \cite{taylor:2008} uses expectiles to efficiently estimate the expected shortfall (ES), a coherent and subadditive risk measure. Its estimation would normally base on a small subset of the available sample. In contrast, the estimate based on expectiles contains all observations. Recent results by \cite{Zie2013} also show that expectiles themselves are a coherent and elicitable risk measure while quantiles are not coherent.
\end{itemize}
In summary, albeit having a different (and may be less intuitive) interpretation than quantiles, expectiles are probably not more difficult to interpret than a variance.

\subsection{Asymmetric Normal Distribution}
To make expectile regression accessible in a Bayesian formulation, we require the specification of an auxiliary response distribution that yields a likelihood that is equivalent to the optimisation criterion~\eqref{kneib:eq:expectiles}. For Bayesian quantile regression, this can be formalised based on the asymmetric Laplace distribution, see for example \cite{Yue}, \cite{Lum} or \cite{YuMoy2001}. For expectile regression, the analogous distribution is an asymmetric normal distribution
\[
 y_i \sim \mathrm{AN}(\eta_i, \sigma^2, \tau)
\]
with density
\[
 p(y_i) = \frac{2}{\sqrt{\sigma^2\pi}}\left(\sqrt{\frac{1}{1-\tau}}+\sqrt{\frac{1}{\tau}}\right)\exp\left(-\frac{1}{2\sigma^2}w_{\tau}(y_i,\eta_{i\tau})(y_i-\eta_{i\tau})^2\right),
\]
expectation
\[
\dsE y_i = \eta_{i,\tau} + \frac{\sigma}{\sqrt{\tau} + \sqrt{1-\tau}} \left( \frac{1-2\tau}{\sqrt{\pi\tau(1-\tau)}} \right)
\]
and variance
\[
\Var(y_i) = \frac{\sigma^2}{\sqrt{\tau}+\sqrt{1-\tau}} \left[ \frac12 \left( \frac{\sqrt{1-\tau}}{\tau} + \frac{\sqrt{\tau}}{1-\tau} \right) - \frac{1}{\sqrt{\tau} + \sqrt{1-\tau}} \left( \frac{(1-2\tau)^2}{\pi\tau(1-\tau)} \right) \right].
\]
Among the several available generalisations of a normal distribution (e.g. the skew normal), maximising the likelihood arising from this distributional specification is then equivalent to minimising~\eqref{kneib:eq:expectiles}, as the logarithmic kernel of the distribution is the same (but negative) argument.

\subsection{Semiparametric Regression}

Instead of only considering linear regression specifications, we are interested in applying expectile regression in the context of general semiparametric regression models with predictor
\[
 \eta_i = \beta_0 + \sum_{j=1}^pf_j(\boldsymbol{z}_i)
\]
where we suppress the index $\tau$ for notational simplicity, $\beta_0$ is an intercept representing the overall level of the predictor, and the functions $f_j(\boldsymbol{z}_i)$ reflect different types of regression effects depending on subsets of the covariate vector $\boldsymbol{z}_i$. For the regression functions $f_j$, we make the following assumptions:
\begin{itemize}
\item The functions $f_j$ are approximated in terms of basis function representations
\[
 f_j(\boldsymbol{z})=\sum_{k=1}^K\beta_{jk}B_k(\boldsymbol{z})
\]
where $B_k(\boldsymbol{z})$ are the basis functions and $\beta_{jk}$ denote the corresponding basis coefficients.
\item The prior for the vector of basis coefficients $\boldsymbol{\beta}_j=(\beta_{j1},\ldots,\beta_{jK})'$ is a multivariate normal distribution with density
\[
 p(\boldsymbol{\beta}_j|\delta_j^2)\propto\exp\left(-\frac{1}{2\delta_j^2}\boldsymbol{\beta}_j'\boldsymbol{K}_j\boldsymbol{\beta}_j\right)
\]
where the precision matrix $\boldsymbol{K}_j$ represents different types of structural assumptions about the function $f_j$ such as smoothness. Note that the prior may be partially improper if the precision matrix $\boldsymbol{K}_j$ is not of full rank.
\end{itemize}

This framework covers, among others, individual-specific random effects, interaction surfaces based on either radial basis functions or tensor product splines, and varying coefficient terms as special cases and therefore provides a convenient generalisation of additive (mixed) models, see \cite{FahKneLan04}. In this paper we will focus on linear effects, penalised splines and Markov random fields, as those will be used in the analysis of childhood malnutrition. For each of those effects we use an appropriate design matrix $\boldsymbol{B}_j$, which renders possible to estimate the predictor $\boldsymbol{\eta}$ as the sum over products $\boldsymbol{B}_j\boldsymbol{\beta}_j$.

\begin{itemize}
\item Linear effects: design matrix $\boldsymbol{B}_j=\boldsymbol{X}$ the data matrix, no penalty matrix at all.
\item Penalised splines: design matrix $\boldsymbol{B}_j$ consists of B-spline basis functions,
precision matrix $\boldsymbol{K} = \boldsymbol{D}_k^{\top}\boldsymbol{D}_k$  with $\boldsymbol{D}_k$ being the matrix of differences of $k$th order.
\item Markov random fields: design matrix $\boldsymbol{B}_j$ consists of the indicator function for the regions, precision matrix $\boldsymbol{K}$ is the neighbouring or adjacency matrix.
\end{itemize}

\subsection{Bayesian Inference}

We complete the Bayesian specification by assuming inverse gamma priors for the error variance and the smoothing variances, i.e.
\[
 \sigma^2\sim\mathrm{IG}(a_0,b_0)\qquad \delta_j^2\sim\mathrm{IG}(a_j,b_j).
\]
Given the model specification, this implies that the full conditionals of the variance parameters are also inverse gamma with updated parameters.
In contrast, the full conditionals for the regression coefficients $\boldsymbol{\beta}_j$ are not available in closed form since unfortunately a normal prior in combination with an asymmetric normal observation models does not induce an asymmetric normal full conditional. We therefore construct proposal densities based on the penalised iteratively weighted least squares updates that would have to be performed to compute penalised expectile regression estimates in a frequentist backfitting procedure, i.e.
\[
 \hat{\boldsymbol{\beta}}_{j\tau}^{[t+1]} = (\boldsymbol{B}_j^{\top}\boldsymbol{W}_\tau^{[t]}\boldsymbol{B}_j + \lambda_j\boldsymbol{K}_j)^{-1}\boldsymbol{B}_j^{\top}\boldsymbol{W}_\tau^{[t]}(\boldsymbol{y}-\boldsymbol{\eta}_{-j,\tau}^{[t]}),
\]
where $\boldsymbol{B}_j$ is the design matrix associated with the $j$-th model term, $\boldsymbol{y}$ is the vector of responses, $\boldsymbol{\eta}_{-j,\tau}=\boldsymbol{\eta}_\tau-\boldsymbol{B}_j\boldsymbol{\beta}_j$ is the complete predictor without the $j$th component,  $\boldsymbol{W}_\tau=\mathrm{diag}(w_\tau(y_1,\eta_{1\tau}),\ldots,w_\tau(y_n,\eta_{n\tau}))$ and $\lambda_j=\sigma^2/\delta_j^2$ is the smoothing parameter obtained as the ratio of error scale parameter and smoothing variance.
More precisely, we propose a new state for $\boldsymbol{\beta}_j$ from the normal distribution $N(\boldsymbol{\mu}_j,\boldsymbol{\Sigma}_j)$ with expectation and covariance matrix given by
\[
 \boldsymbol{\mu}_j=\boldsymbol{\Sigma}_j\boldsymbol{B}_j^{\top}\boldsymbol{W}_\tau(\boldsymbol{y}-\boldsymbol{\eta}_{-j,\tau}) \quad\mbox{and}\quad\boldsymbol{\Sigma}_j=(\boldsymbol{B}_j^{\top}\boldsymbol{W}_\tau\boldsymbol{B}_j + \lambda_j\boldsymbol{K}_j)^{-1}.
\]

This framework also allows to extend the algorithm to shrinkage or variable selection algorithms such as the Bayesian LASSO which has already been done in Bayesian quantile regression \citep[see][]{Waldmann}.

\section{Simulations}
Since we need a misspecified likelihood for our estimations, we aim to show that the resulting estimated expectiles are nevertheless valuable. We therefore conduct simulation studies comparing Bayesian expectile estimates with least squares and boosting estimates \citep{Sobotka:2010} in order to quantify the performance of the procedures. 
The estimates are rated according to the true expectiles of the error distribution. 

First, we evaluate the posterior mean as a point estimate and afterwards we explore coverage rates and the widths of the credible intervals.

\subsection{Point Estimates}

\subsubsection{Design}
To start the evaluation of the Bayesian expectiles and for comparison with existing alternatives, we generate two covariates, $X_1 \sim B(1,0.5)$ and $Z_2 \sim U(0,3)$ in sample sizes of $n=100,500$. Next, the random errors $\varepsilon$ are drawn from an A) $N(0,0.5 z_2^2)$, B) $\textnormal{Exp}\left(\frac{1}{z_2}\right)$ or C) $t(2)$ distribution. Together they comprise data for two simple semiparametric models in the following way:
\begin{eqnarray*}
  (\textrm{M1}) & \yvec =& 2 \xvec_1 + 5 \exp(-0.5 \zvec_2^2) + \varepsilonvec \\
  \textrm{(M2)} & \yvec =& 2 \xvec_1 + 5 \sin(2 \zvec_2) + \varepsilonvec.
\end{eqnarray*}
Hence, we have a challenging homoscedastic scenario (C) with infinite variance and two heteroscedastic scenarios, one of them with skewed errors (B). For each of the combinations of sample size, error distribution and model formula we generate 100 replications. The data are then analysed for expectiles with asymmetries $\tau \in \{0.02,0.05,0.1, \allowbreak 0.2,0.5,0.8, \allowbreak 0.9,0.95,0.98\}$. We estimate the Bayesian expectiles with overall 35000 MCMC iterations, where 5000 are burn-in and we use a thinning of 30. This leaves us with a sample of 1000 observations from the posterior. This method is compared with a least asymmetrically weighted squares (LAWS) estimate and an estimate obtained with the use of component-wise functional gradient boosting, both as presented in \cite{Sobotka:2010}. The smoothing parameter in LAWS estimation is optimised with an asymmetric cross-validation criterion, for boosting the optimal stopping iteration from 1000 initial boosting iterations is also determined via cross-validation. For all algorithms, the nonlinear effect is estimated using a cubic B-spline basis with 20 inner knots and second order difference penalty. The methods are taken from the software package expectreg \citep{expectreg} available for R \citep{CRAN}.

\subsubsection{Results}

\begin{figure}[ht!]
  \subfloat[ $N(0,0.25z_2^2)$-error]
    {\includegraphics[width=60mm]{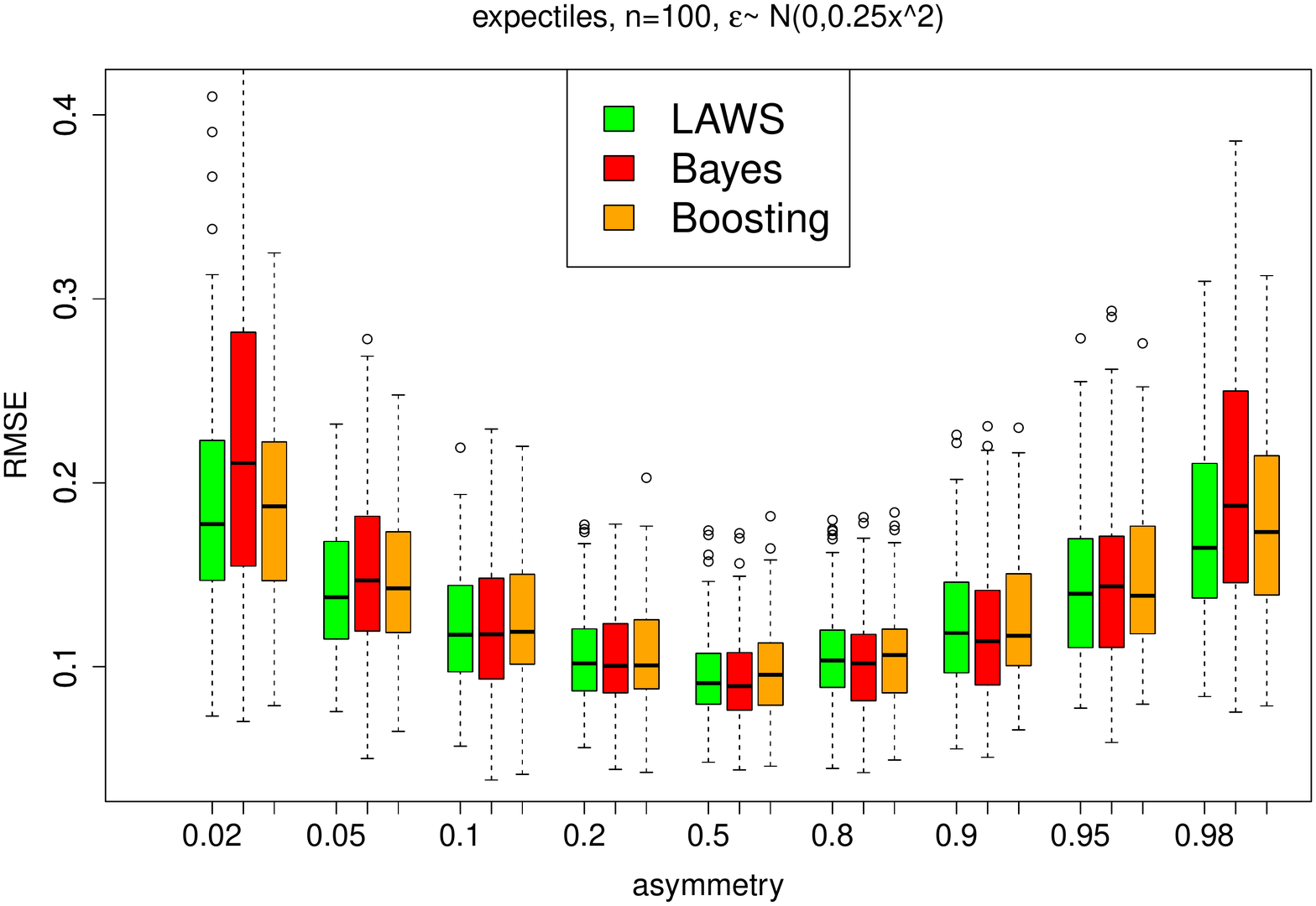} }
  \
  \subfloat[ $\textnormal{Exp}(\frac{1}{z_2})$-error]
    {\includegraphics[width=60mm]{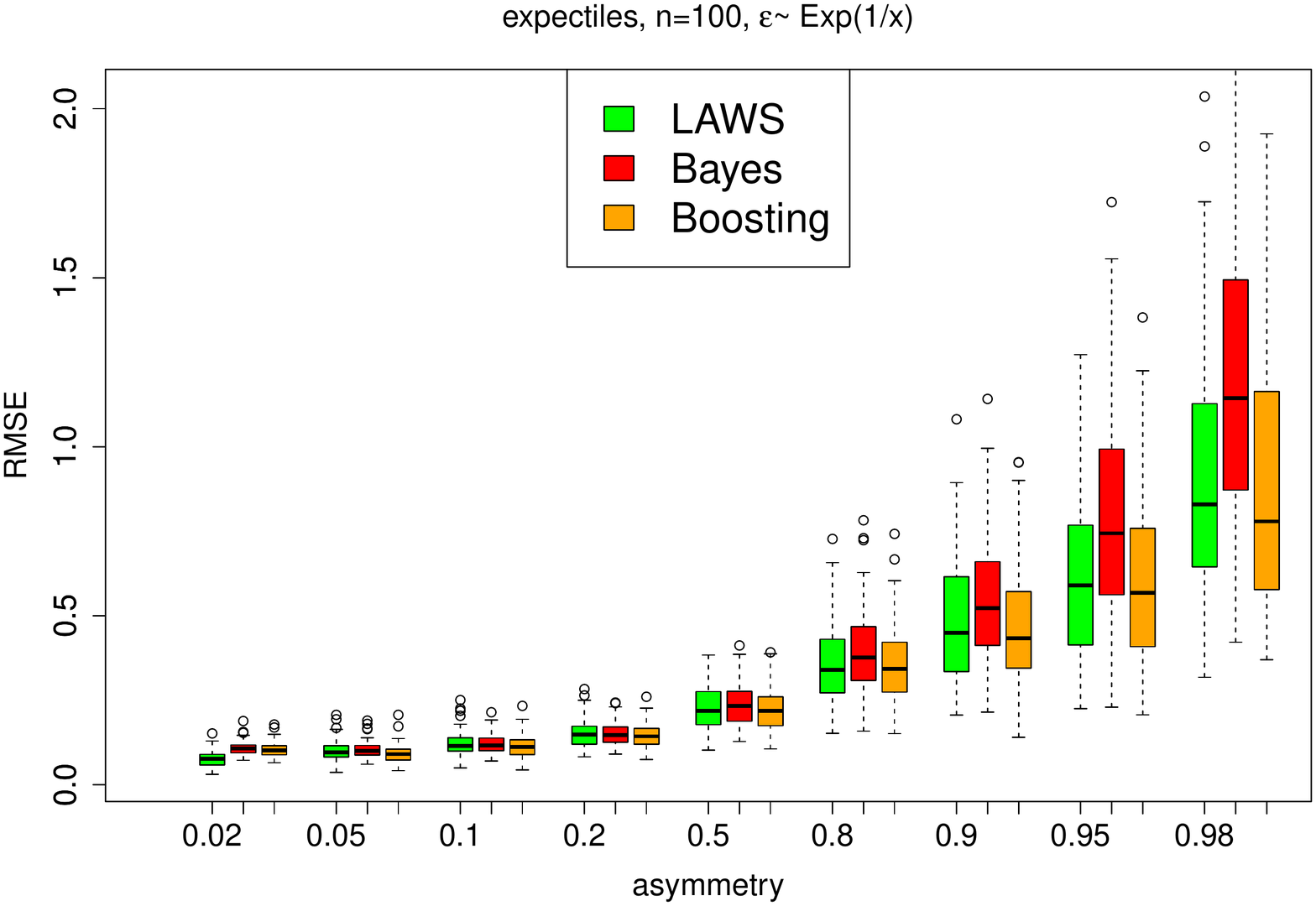} }
  \
  \subfloat[ $t(2)$-error]
    {\includegraphics[width=60mm]{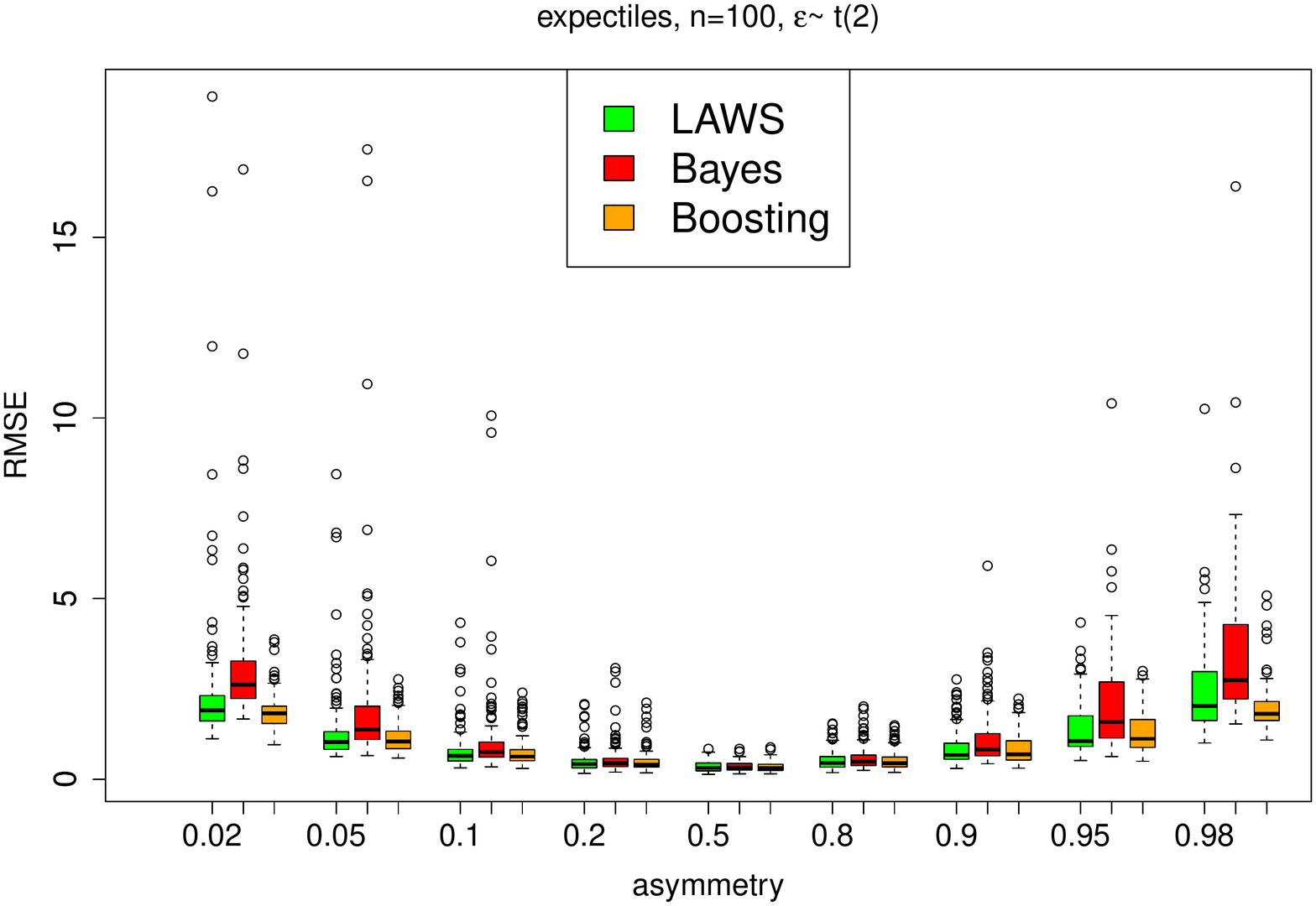} }
  \caption{RMSE for $n=500$, the three different errrors, separately for each expectile. Boxplots created from 100 replications.}\label{plot:sim:mse}
\end{figure}

The quality of the results will be measured in terms of a root mean squared error for the estimated function:
\[
 \RMSE(f_\tau) = \sqrt{(\fvec_\tau-\hat{\fvec}_\tau)'(\fvec_\tau-\hat{\fvec}_\tau)}.
\]
In Figure~\ref{plot:sim:mse} we present the results of the three methods, for each expectile in direct comparison. The results are shown for $n=500$ and exemplary for (M1). The complete results are available as online supplement. Our simulations show that, in terms of RMSE, the methods are quite interchangeable. The posterior means offer the same estimation quality as LAWS and boosting, at least within $\tau \in [0.05,0.95]$. For more extreme expectiles, it seems that the numerical errors in the MCMC  start to become more substantial, i.e. it becomes difficult to draw from the respective multivariate normal distribution with extreme weights. Otherwise, the choice for one of the estimates can be made regarding the outer properties, practicability or just personal habit, now that a Bayesian estimate is available. The choice might also depend on interval estimates rather than point estimates. The former are analysed in the next part of the simulations.

\subsection{Interval Estimates}

\subsubsection{Design}
Similar as in \cite{Waldmann} we compare $95\%$ credibility intervals with frequentist confidence intervals based on an asymptotic normal distribution \citep{SobotkaKau:2012}. Confidence intervals from boosting would be obtained with a computationally demanding nonparametric bootstrap and are therefore omitted. The comparison is made regarding coverage properties and interval widths. For a simpler visualisation we focus on a single nonlinear effect:
\[
  \textnormal{(M3)} \, \yvec = \sin(2(4 \zvec - 2)) + 2 \exp(-16^2 (\zvec - 0.5)^2) + \varepsilonvec.
\]
The covariate is drawn from a $U(0,1)$ distribution,  the error from a $N(0,(0.2+|z-0.5|)^2)$ and an $\textnormal{Exp}\left(\frac{1}{0.2+|z-0.5|}\right)$ distribution. The nonlinear effect then has its highest frequency as well as lowest variance at 0.5 while the variance increases with$z \to 0$ and $z \to 1$. The frequentist asymptotics start to apply from 500 observations and for extreme expectiles, 1000 observations are recommended. Hence, we generate data sets with $n=500,1000$ and in order to properly measure the coverage rate, we generate 1000 replications. The rest of the parameters remain as before.

\subsubsection{Results}

\begin{figure}[ht!]
  \subfloat[Bayesian example analysis]
    {\includegraphics[width=60mm]{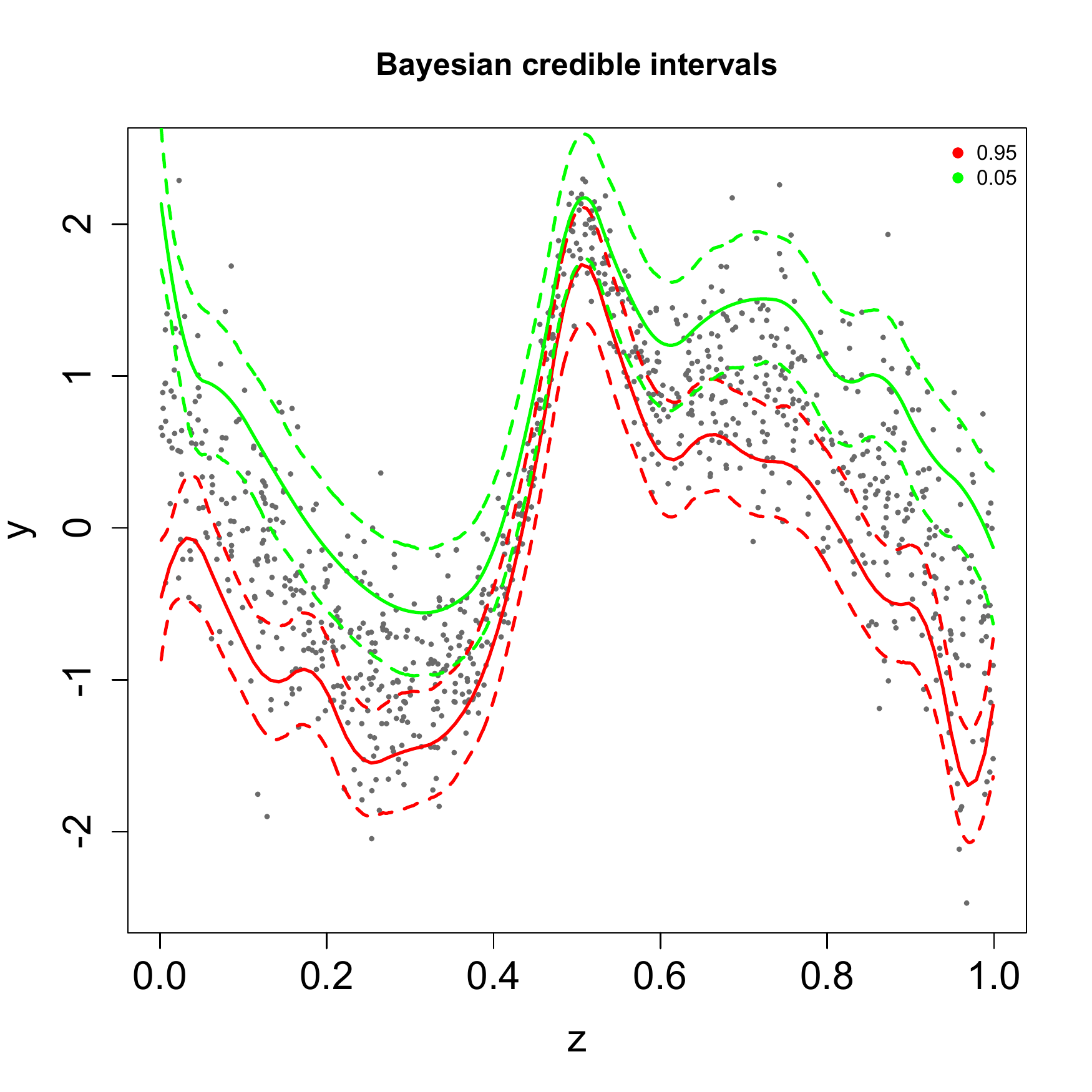} }
  \
  \subfloat[frequentist example analysis]
    {\includegraphics[width=60mm]{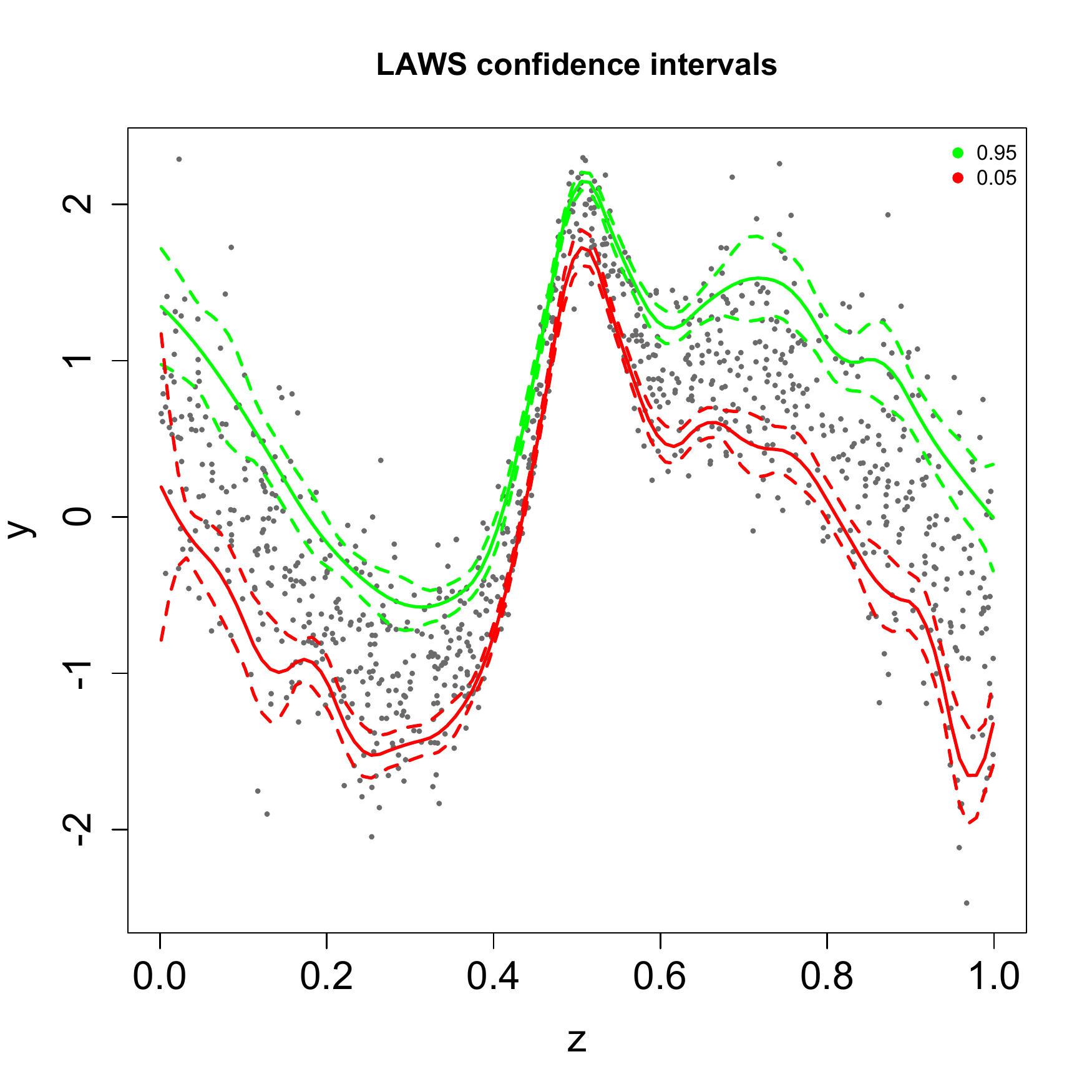} }
  \caption{Examplary estimates and pointwise intervals for $n=1000$, (M3) and normal errors obtained from MCMC and LAWS estimation.}\label{plot:sim:ex}
\end{figure}

\begin{figure}[ht!]
  \subfloat[coverage for $\tau = 0.05$]
    {\includegraphics[width=60mm]{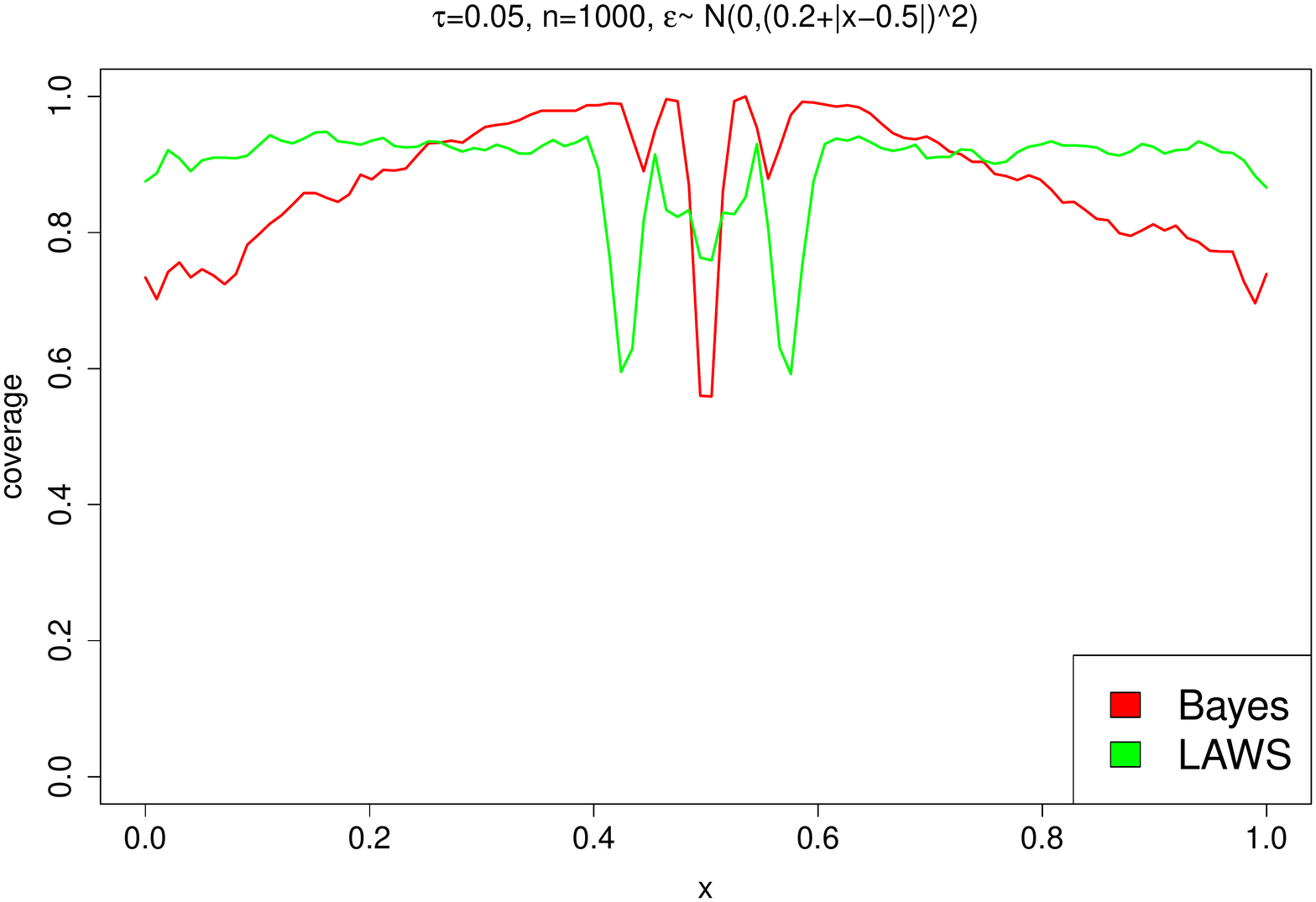} }
  \
  \subfloat[min and max width for $\tau = 0.05$]
    {\includegraphics[width=60mm]{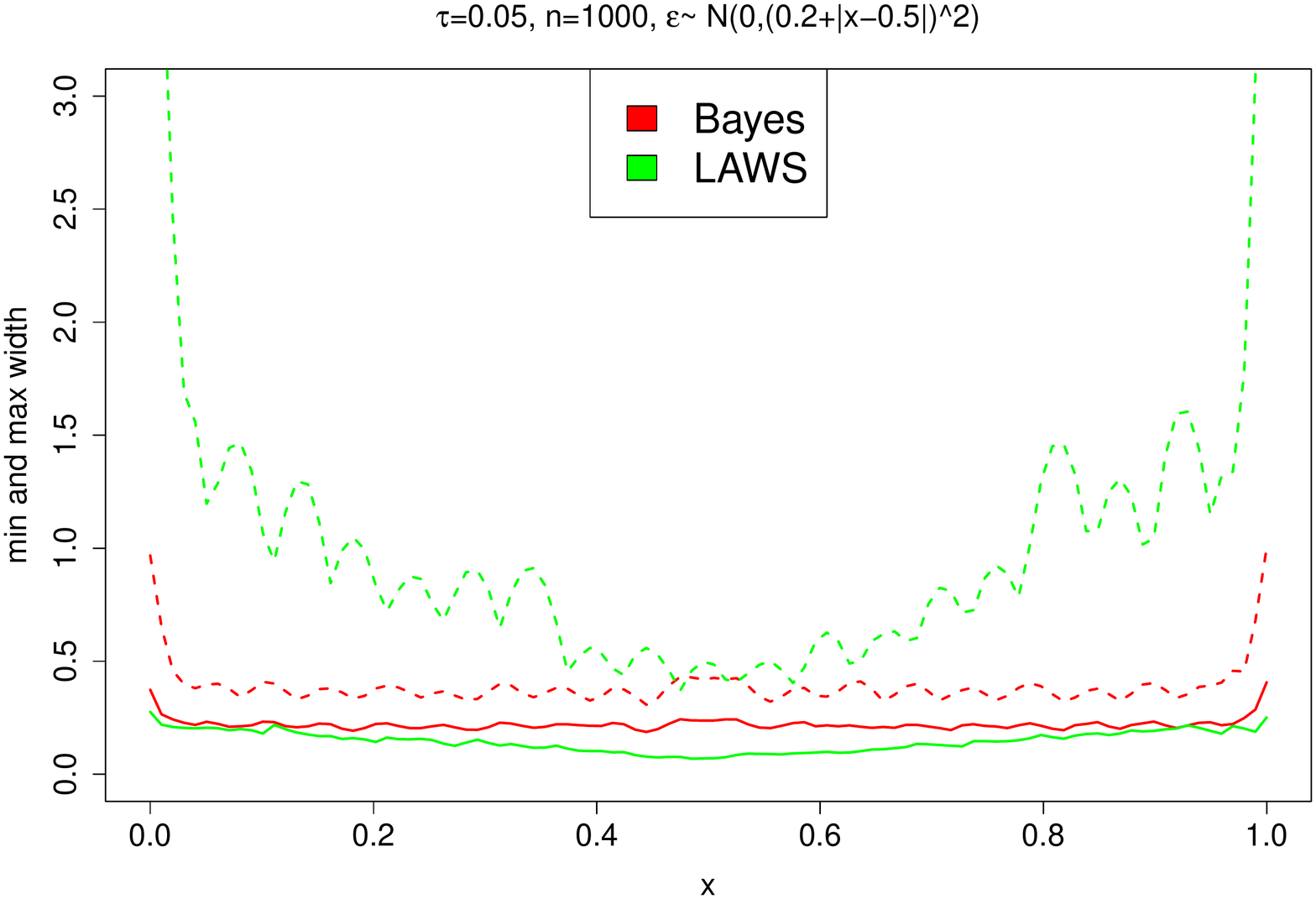} }
  \
  \subfloat[coverage for $\tau = 0.5$]
    {\includegraphics[width=60mm]{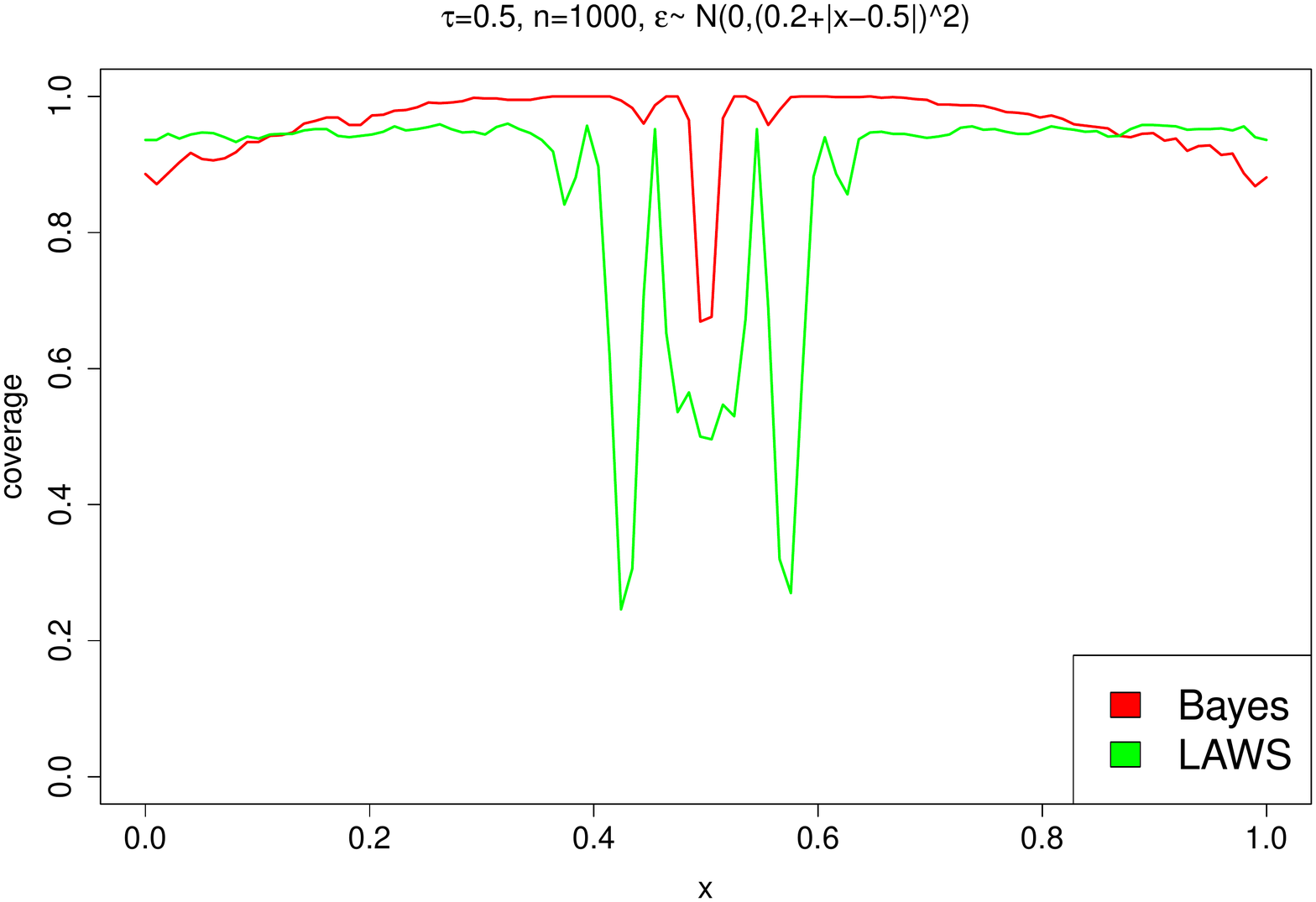} }
  \
  \subfloat[min and max width for $\tau = 0.5$]
    {\includegraphics[width=60mm]{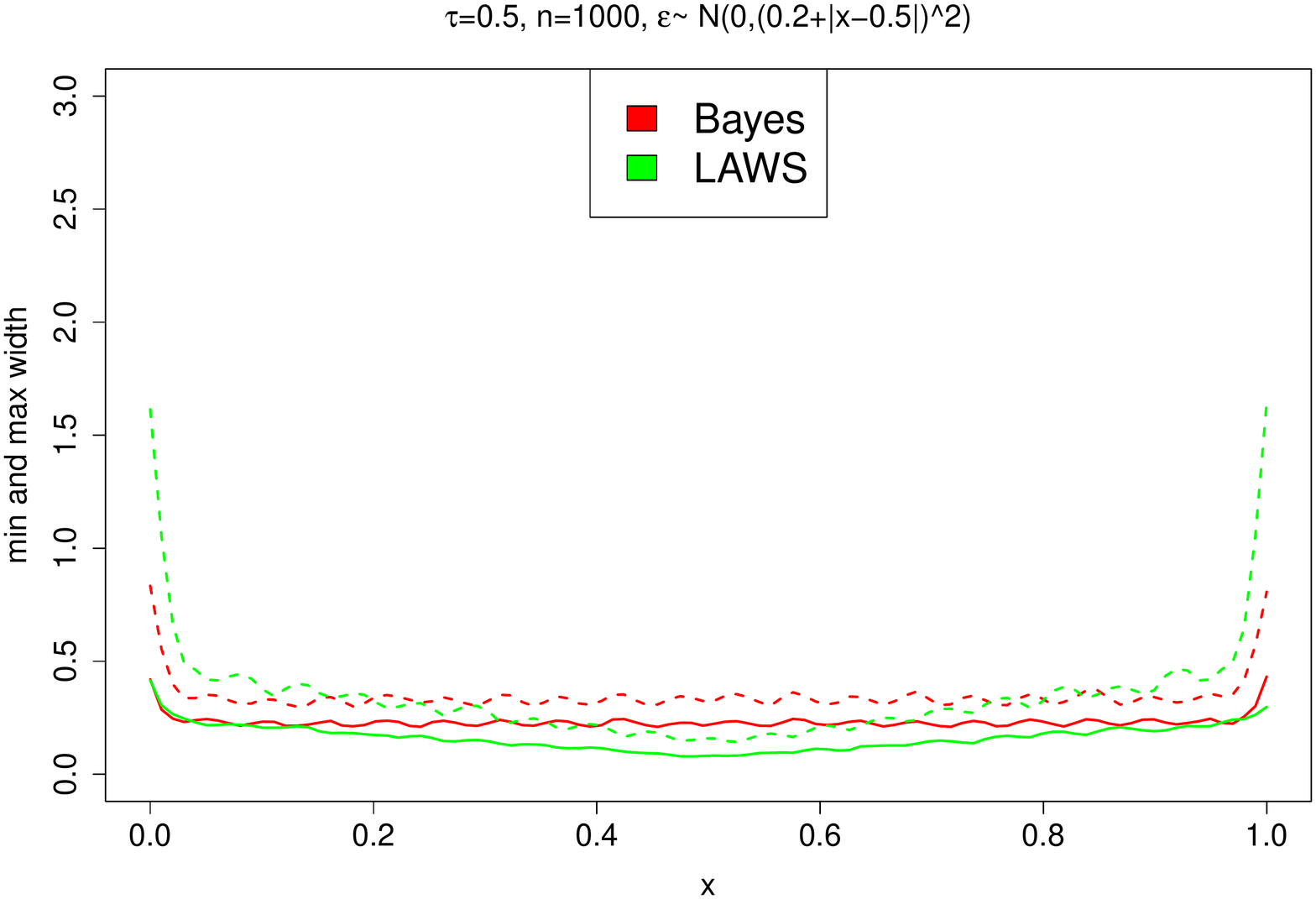} }
   \
  \subfloat[coverage for $\tau = 0.95$]
    {\includegraphics[width=60mm]{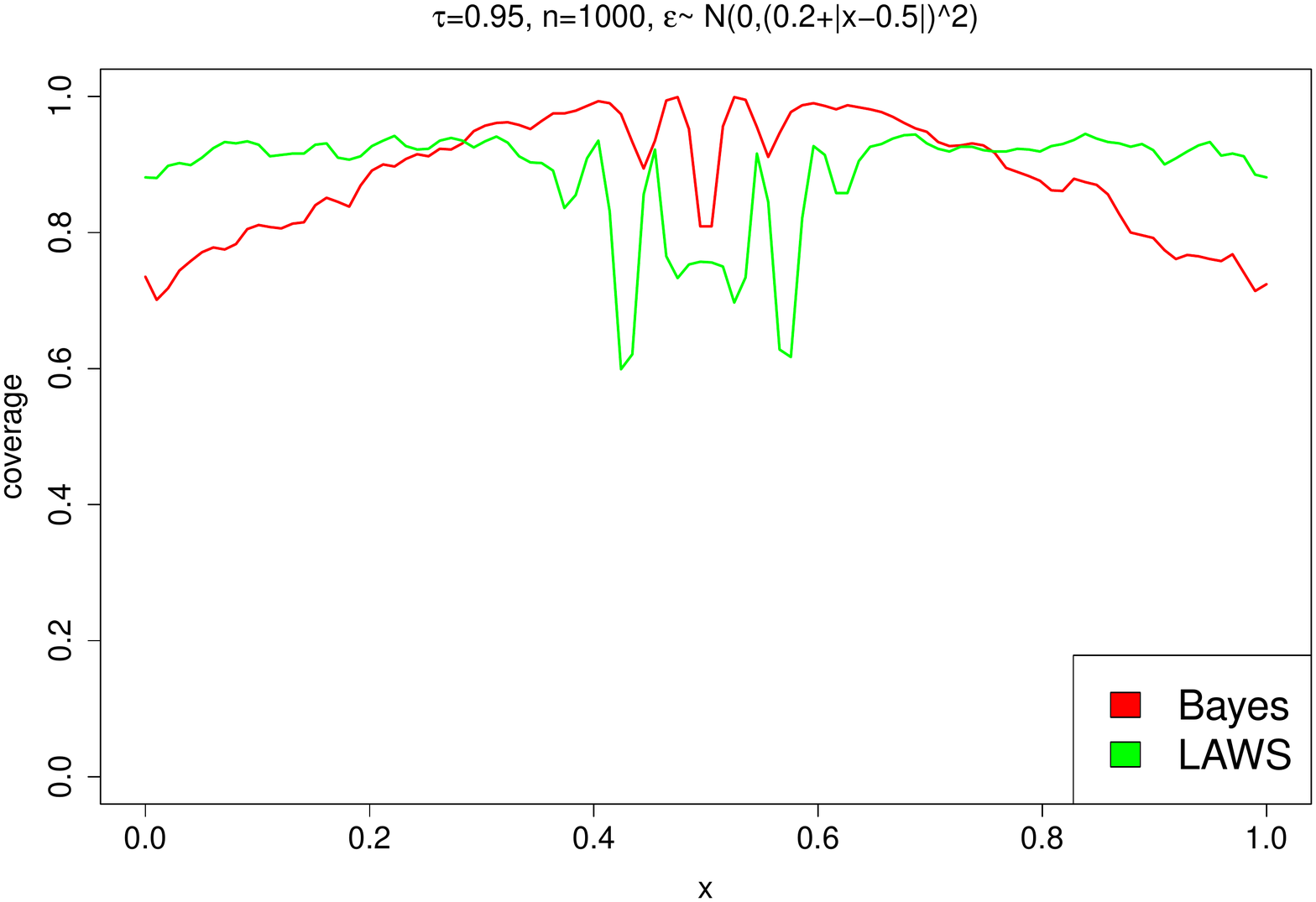} }
   \hfill
  \subfloat[min and max width for $\tau = 0.95$]
    {\includegraphics[width=60mm]{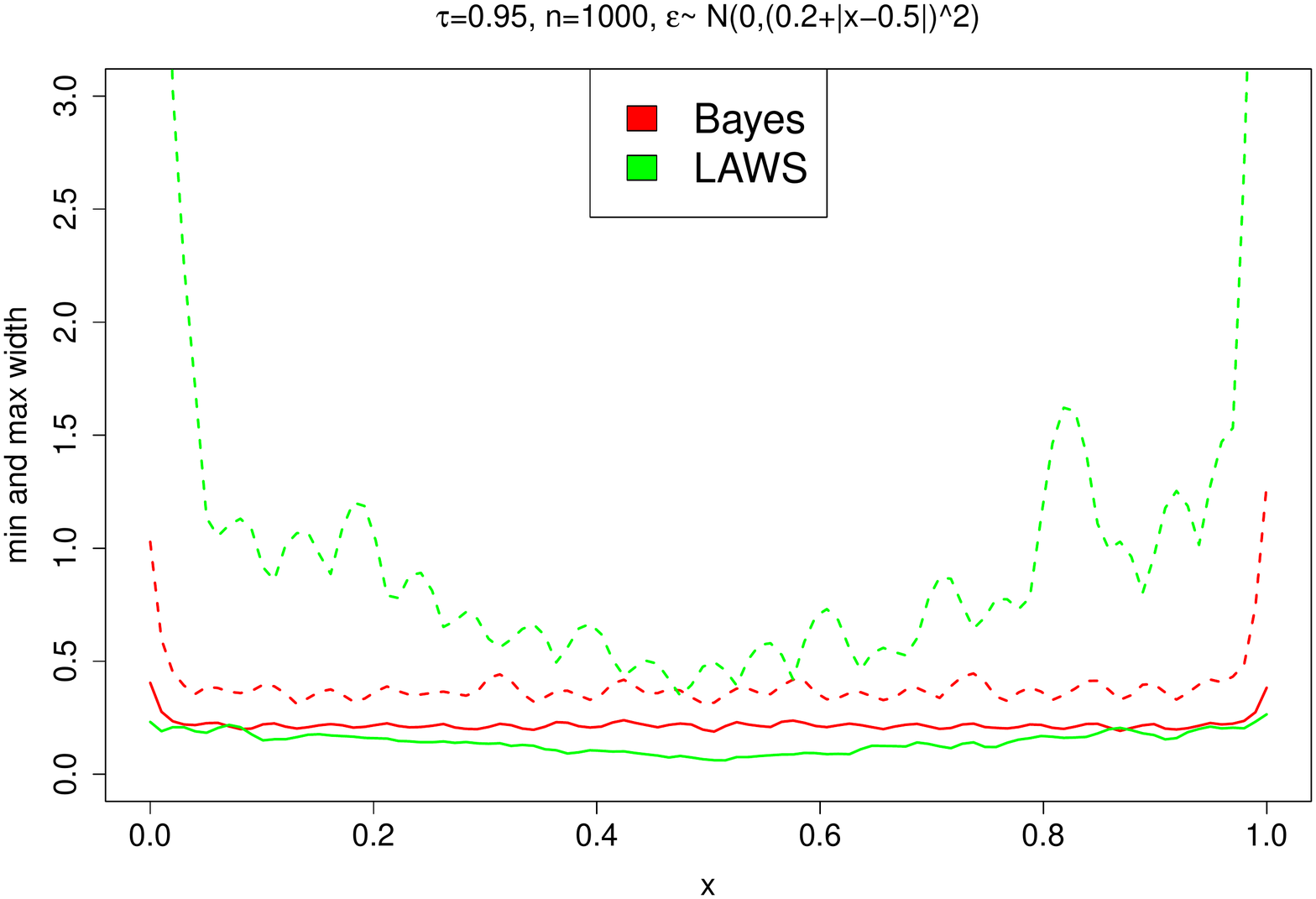} }
  \caption{Coverage rates on the left and interval widths on the right for normal errors and $n=1000$. Minimal interval width given in solid, maximum width in dashed lines.}\label{plot:sim:ki}
\end{figure}

We measure the coverage of the confidence intervals at a given covariate value $ z_{i} $ as
\[
  \widehat{\textnormal{Cover}}(CI(\hat{f}_{j,\tau}(z_{i})) = \frac{1}{1000} \sum_{k=1}^{1000}{\mathds{1}_{\{\hat{f}_{j,\tau}(z_{i}) \in CI(\hat{f}_{j,\tau}^{[k]}(z_{i}))\}}} \, ,
\]
the maximum width of all confidence intervals at all fixed $z_{i}$
\begin{equation*}
  \max \widehat{\textnormal{Width}}(CI(\hat{f}_{j,\tau}(z_{i}))) = \max_k ( \hat{f}_{j,\tau,U}^{[k]}(z_{i}) - \hat{f}_{j,\tau,L}^{[k]}(z_{i}))
\end{equation*}
where $f_U$ and $f_L$ denote the upper and lower ends of the interval estimate. The minimum width is determined in the same way. The evaluations are done on a regular grid of length 100 within the covariate domain. In Figure~\ref{plot:sim:ki} we can see that the coverage of both interval estimates is rather poor at the center of the covariate where the curvature of the generating function is high. This might result from a bias that comes with the addition of the penalty. Otherwise the plots show that the width of the frequentist intervals generally increases with increasing variance in the errors while the credibility intervals remain at the same width over the whole covariate domain. The effect is a better coverage at the center of the covariate and worse coverage for strongly increasing variance regions. This result is especially visible in Figure~\ref{plot:sim:ex} where two estimates and intervals are shown for an exemplary data set. Here we can see that the confidence intervals are much narrower in the center than the credible intervals. Reasons for this behaviour can be found in the misspecified likelihood which is just an auxiliary tool to fit the point estimates and does not describe the data well. Hence, the estimated variance of the fit is constant. The results for exponential errors and for a sample size of 500 are available as online supplement.

\subsection{Simulation roundup}
Overall we can say that boosting is a flexible tool that results in good point estimates, confidence intervals for large data sets with strong heteroscedasticity might be more reliable with a LAWS estimate, but the estimated Bayesian expectiles are as efficient and provide better coverage for small samples.
\section{Example}\label{sec:ex}
A data set consisting of 5389 observations of children was obtained from the Demographic Health Surveys (DHS, \texttt{www.measuredhs.com}). The study was conducted in Tanzania in 1992. It contains information on weight, height, sex and age of the children themselves, information about the parents - namely the mother - such as her BMI at the birth of the child, her educational background (in four categories) as well as her current employment status (either employed or unemployed) and the information on the residence. The latter actually splits into two: the categorical variable on the surrounding (urban or rural) and the spatial variable, indicating the province mother and child are living in. Chronic malnutrition leads to {\it stunting} (insufficient height for age) which will be used as measure for the extent of undernourishment. The height of the children is compared to a reference population of supposedly healthy children of the same age in a so called {\it z-score}: $z_i = (\textit{AI}_i - \textit{MAI})/\sigma$. In this formula $\textit{AI}_i$ stands for the stunting index of child $i$, $\textit{MAI}$ the median stunting value in the reference population and $\sigma$ for the standard deviation of stunting in the reference population. The mean value in our data set is $-177.9$, the standard deviation $142.24$, 90\% of the children have a z-score lower than zero and the 95\%-quantile reaches from $-455.00$ to $108.05$. As explained in the introduction, we use a model incorporating the continuous covariates nonlinearly, the categorical variables linearly and the province as a spatial effect (see equation(\ref{mod1})). The estimation was executed as described in Section~\ref{sec:bayes}: the nonlinear effects are modeled with Bayesian $P$-splines, the spatial effect with a Markov random fields. The model was estimated for the $\tau = 0.05, 0.1, 0.2, 0.5, 0.8, 0.9$  and $0.95$-expectiles. 

The results for the linear effects are displayed in Table~\ref{tab:app:linear}. Effects, for which the 95\% posterior interval does not contain zero, are printed in boldface. For reasons of clarity only five different expectiles are displayed, the rest behaves analogously. Note that for the covariate {\it work} the sign of the effect changes over the expectiles. This means that the impact of the employment of the mother is positive in the lower parts of the conditional distribution, whereas it has a negativ impact in the middle to higher ends. A similar effect can be seen for the impact of {\it secondary school} in comparison to{\it no education at all}. The positive effect of the variable {\it rural} is no surprise, as the proximity to the farms and the traditional higher family bonding is of high importance for the adequate supply. The negative effect of sex simply displays the fact, that boys of this age are generaly less tall than girls.

\begin{figure}[ht!]
  \subfloat[Nonlinear effect for BMI of mother at birth]
    {\includegraphics[width=70mm]{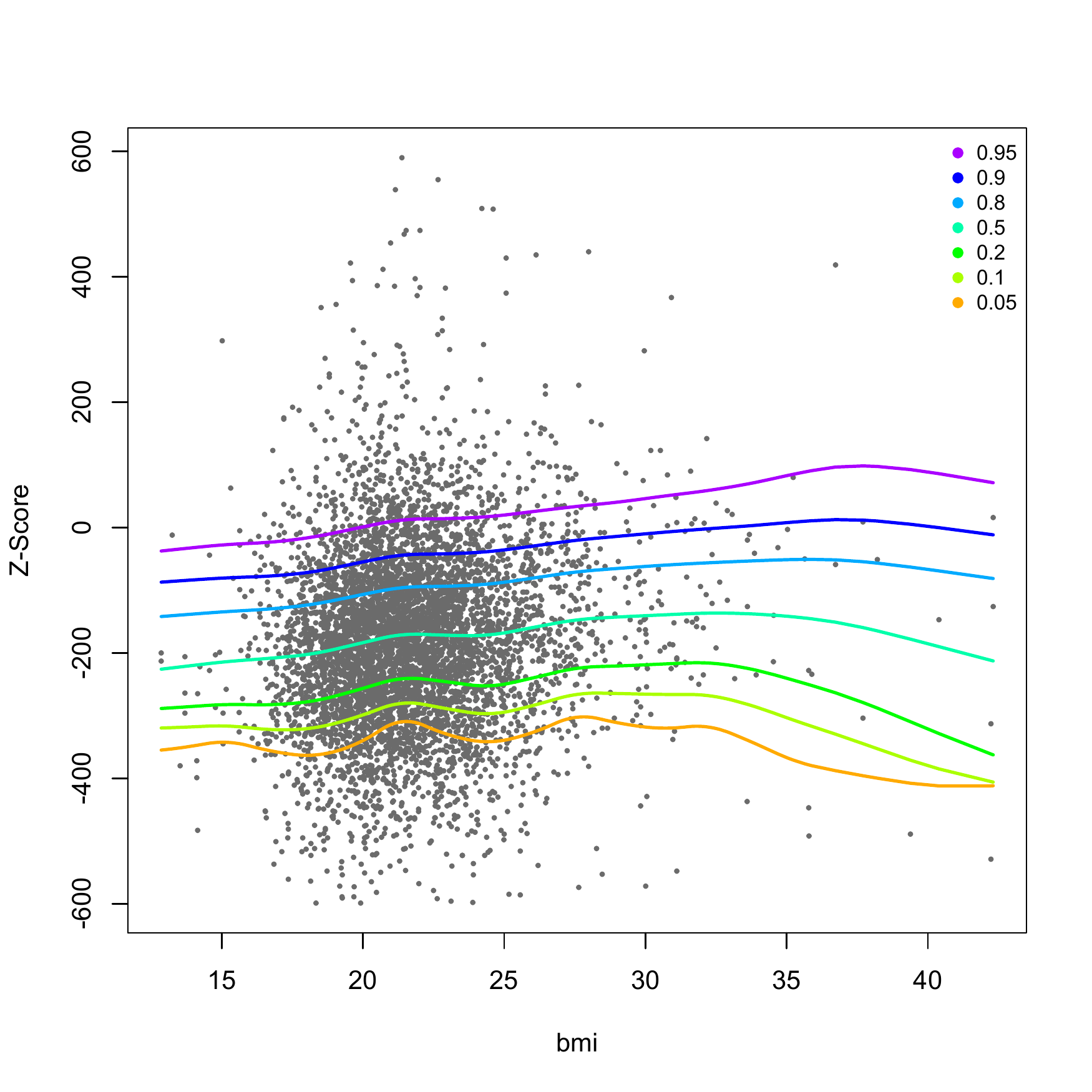} }
  \
  \subfloat[Nonlinear effect for the age of the child in months]
    {\includegraphics[width=70mm]{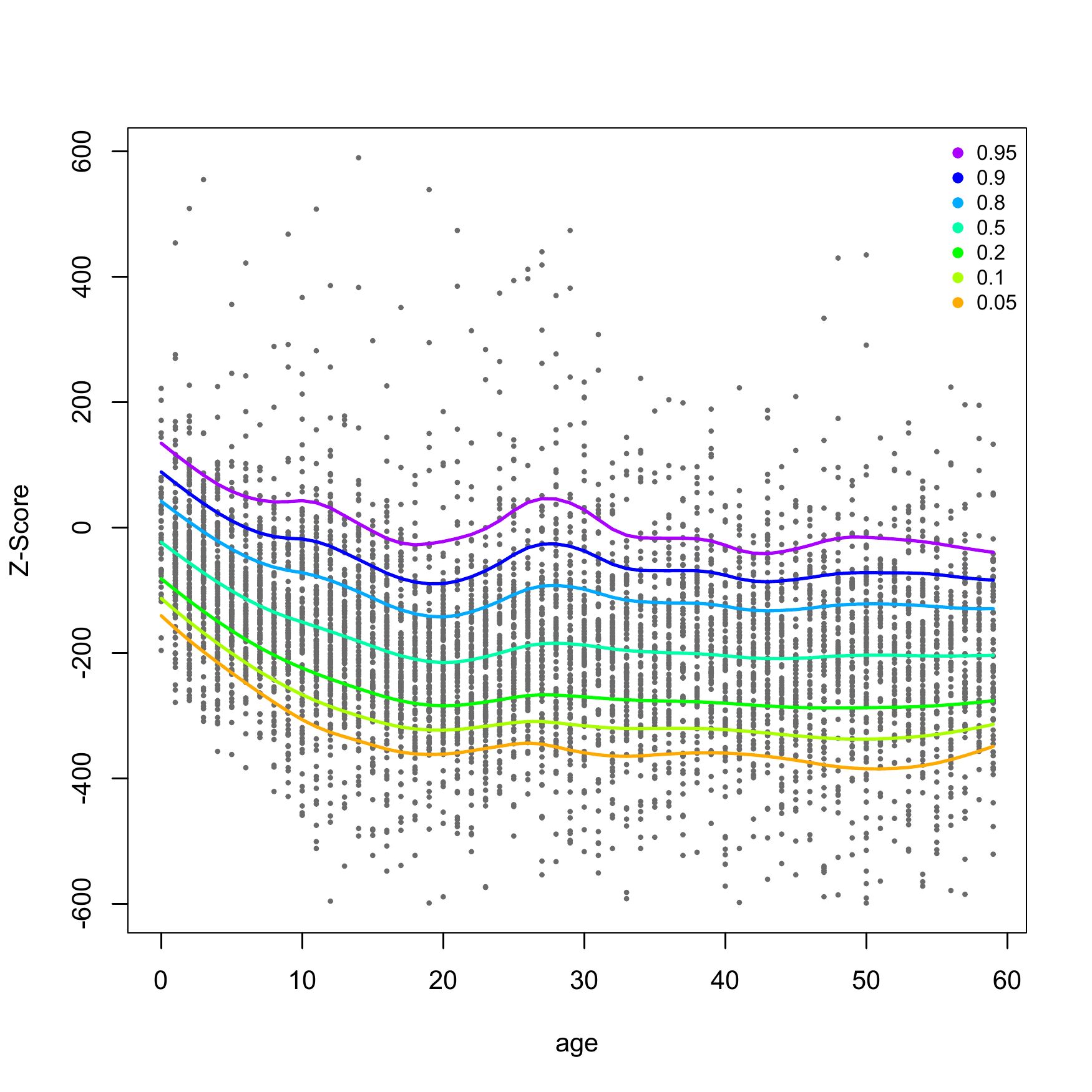} }
  \caption{Estimated nonlinear effects for the childhood malnutrition data. Results for expectiles from 0.05 to 0.95 shown.}\label{plot:app:nonlinear}
\end{figure}

Nonlinear effects are displayed in Figure~\ref{plot:app:nonlinear}. The results are very close to those from \cite{kandala}, where the data set was analysed in a mean regression setting. There are small differences between the expectiles, but in general the effects are stable over the whole distribution.

\begin{figure}[ht!]
  \subfloat[0.05-expectile]
    {\includegraphics[width=70mm]{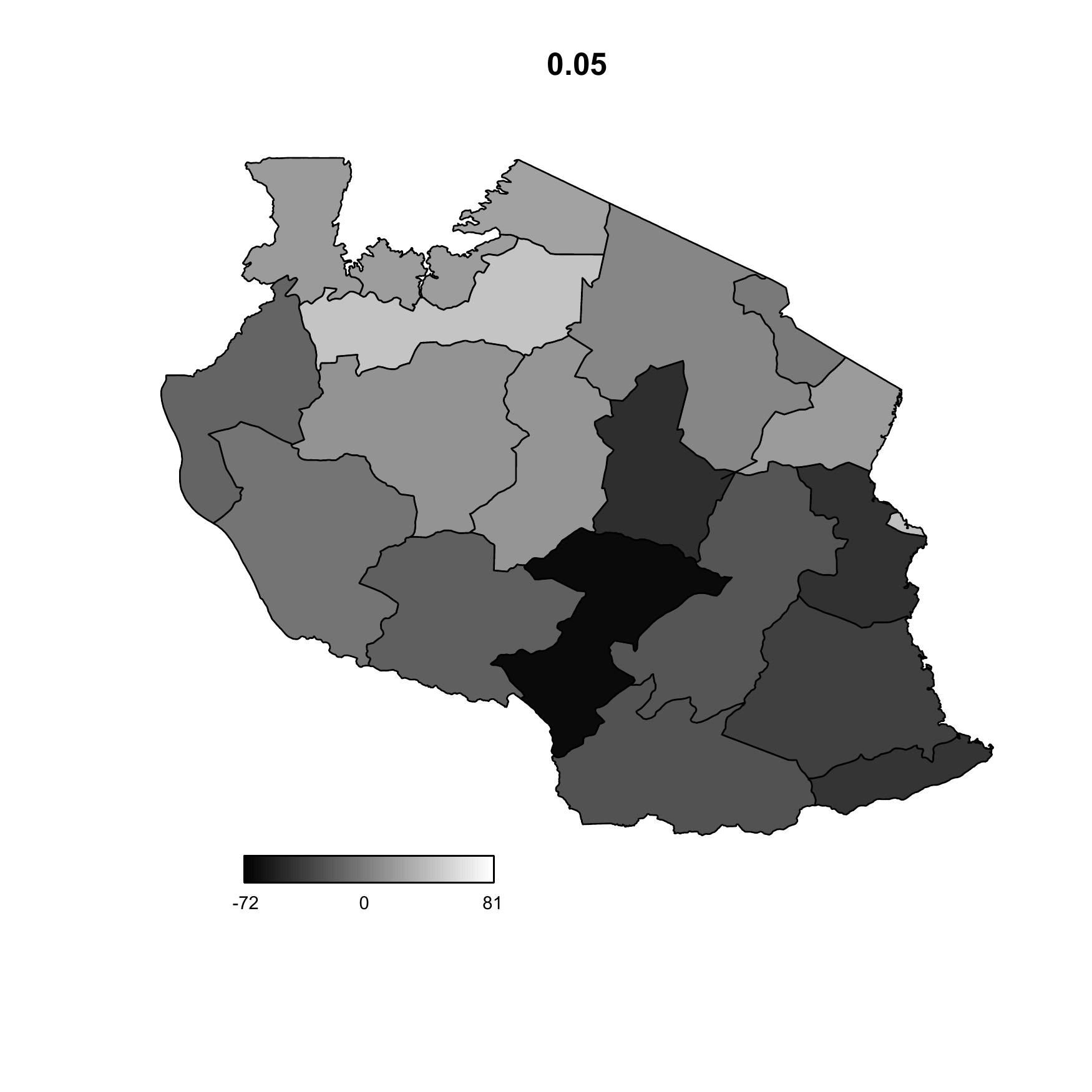} }
  \
  \subfloat[0.2-expectile]
    {\includegraphics[width=70mm]{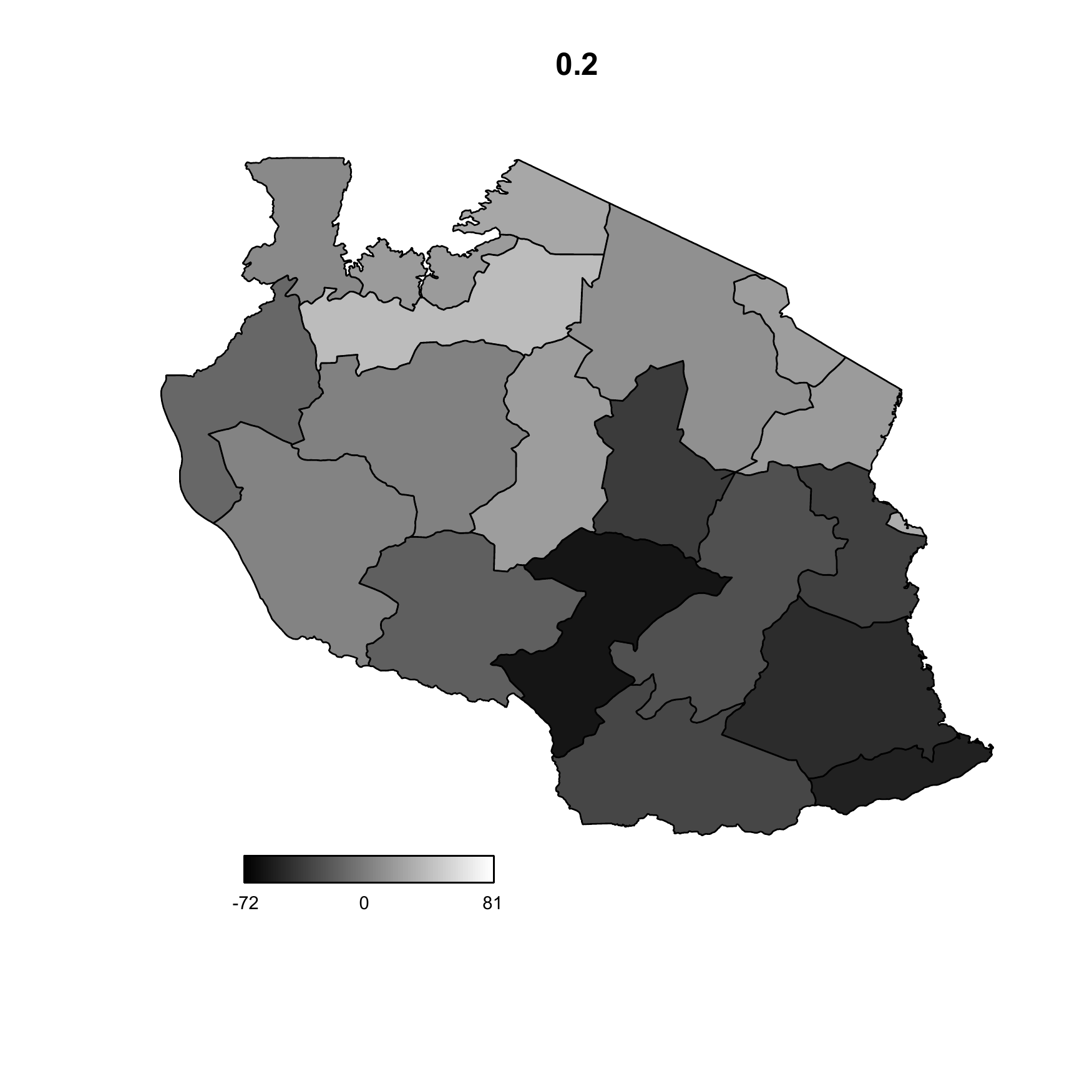} }
  \
  \subfloat[0.8-expectile]
    {\includegraphics[width=70mm]{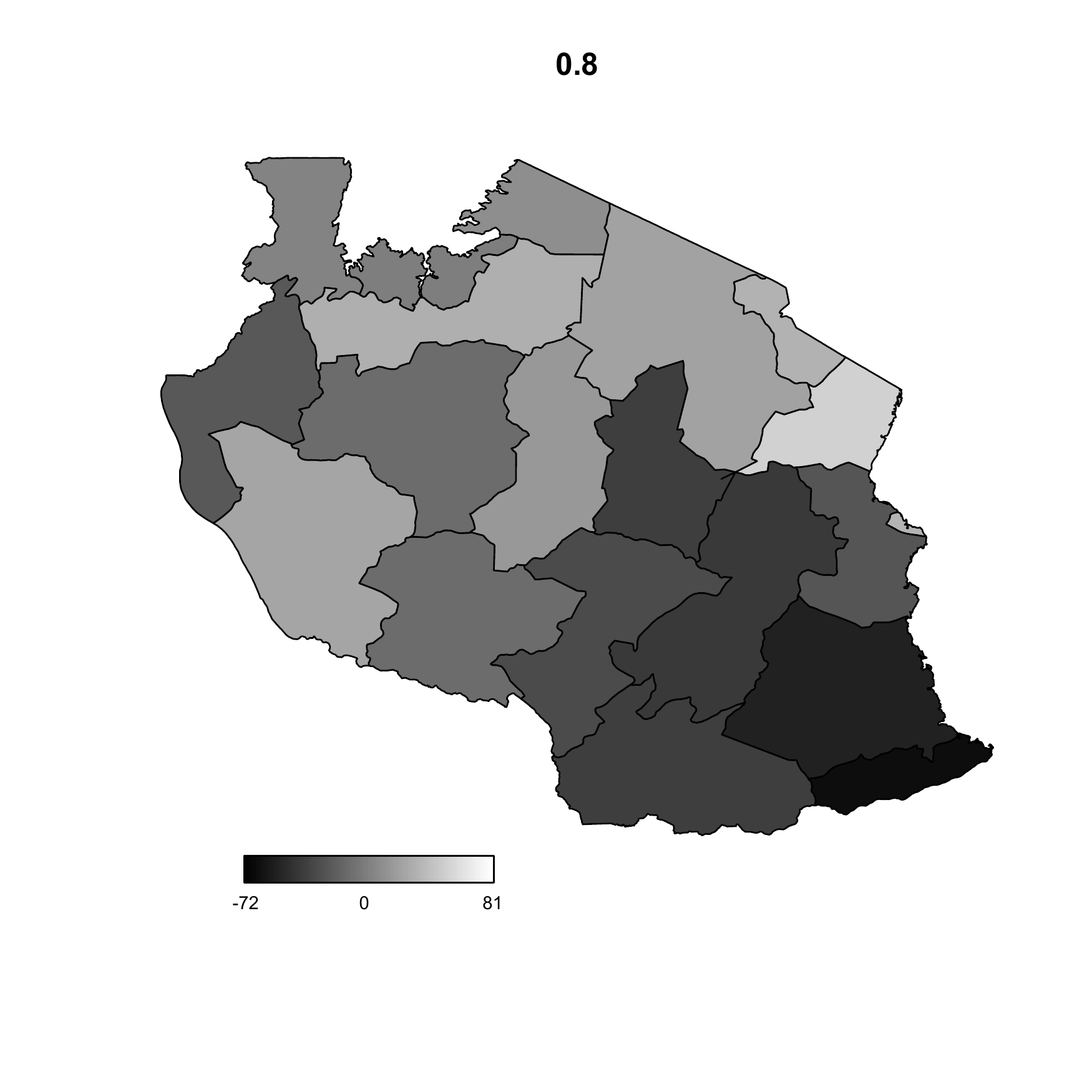} }
  \hfill
  \subfloat[0.95-expectile]
    {\includegraphics[width=70mm]{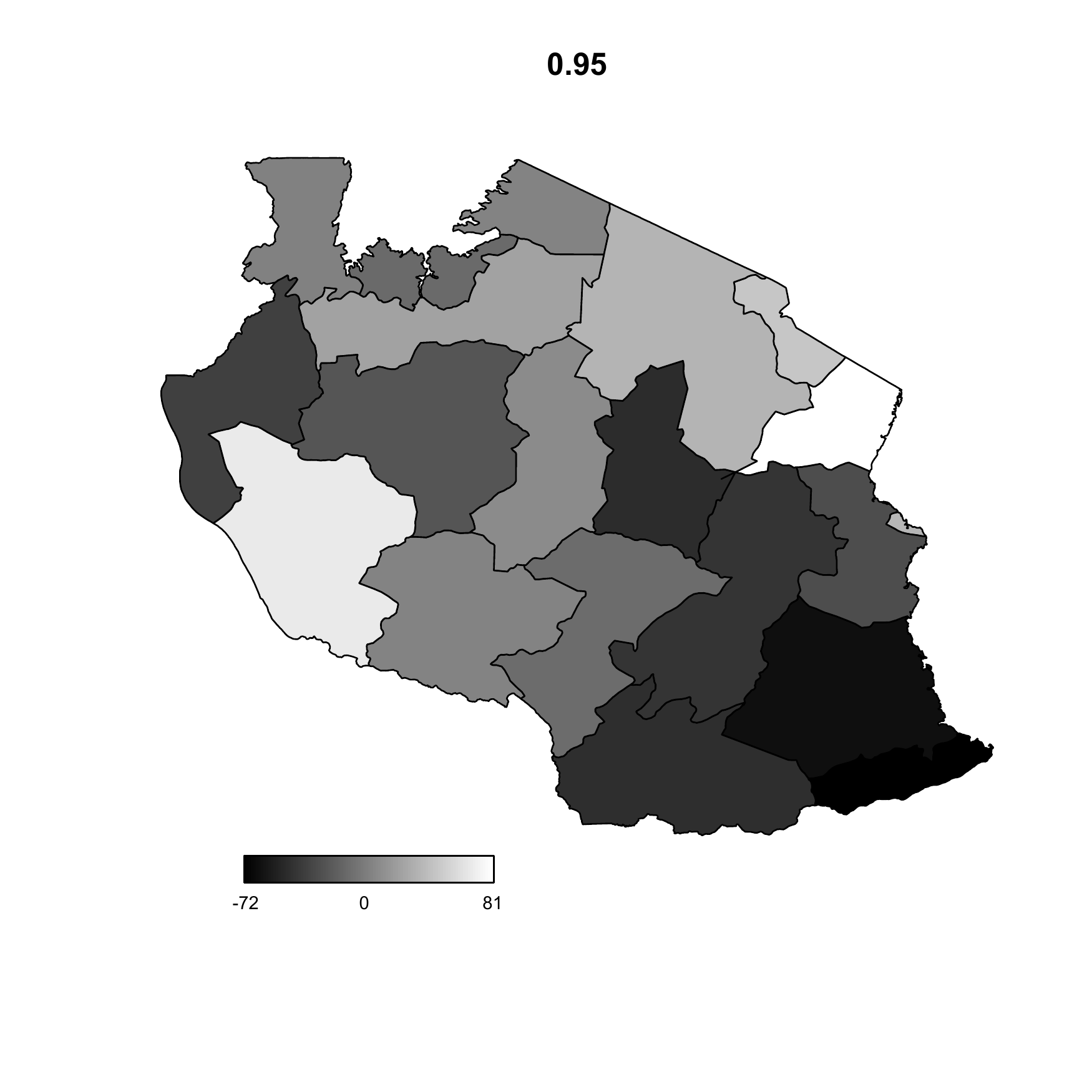} }
  \caption{Estimated spatial effects for the childhood malnutrition data provided in a map of Tanzania.}\label{plot:app:spatial}
\end{figure}

For the spatial effects see Figure~\ref{plot:app:spatial}. The effect of the capital Dar es Salaam in the east of the country is positive over all expectiles, which is in contradiction to the negative effect of the variable {\it urban} in comparison to {\it rural}. Thus we conclude, that the effect of the capital as being 	better supplied than the rest of the country voids this effect. Another fact worth mentioning is the positive effect of the south west on the higher expectile. This region neighbours Lake Tanganyika and is known for its fertility.  








\begin{table*}[h!]
\begin{center}

\begin{tabular}{lrrrr} \hline
Variable / $\tau$ & 0.05 &  0.2  & 0.8 &  0.95  \\ \hline
mother's work                   &  {\bf 9.18} &  1.21 & {\bf -11.71} &    {\bf -25.63} \\
          \small \it reference: ``unemployed''     & \small \it (3.52,15.61) &\small \it (-4.29,6.95) &\small \it (-18.42,-4.87) &\small \it (-33.20,-18.12) \\ mother's education:  & \multicolumn{3}{c}{\small \it reference: ``no education''} \\
 \it ``primary school''            &   -3.15 & -0.65 & {\bf -8.02} &   {\bf  -18.14} \\
           \small \it      &\small \it (-10.45,4.02) &\small \it (-7.44,6.17) &\small \it (-15.64,-0.76) &\small \it (-27.15,-8.81) \\
\it `` secondary school''            &  {\bf 20.48} & {\bf 15.57} & 3.31 &   {\bf -9.71} \\
           \small \it      &\small \it (14.42,26.48) &\small \it (10.10,21.03) &\small \it (-3.19,9.53) &\small \it (-17.49,-1.98) \\
 \it ``higher education''            &  {\bf 60.43} & {\bf 56.94} & {\bf 61.67} &  {\bf  75.13} \\
           \small \it      &\small \it (40.39,79.93) &\small \it (38.65,75.52) &\small \it (40.28,83.07) &\small \it (50.34,101.99) \\
 mother's residence         &   {\bf 24.55} &  {\bf 26.84} & {\bf 18.01} &   7.09 \\
           \small \it reference: ``urban''      &\small \it (15.22,33.71) &\small \it (19.34,33.98) &\small \it (8.96,27.31) &\small \it (-3.06,16.89) \\
 child's sex       &    {\bf -11.54} & {\bf -11.77} &  {\bf -8.25} &  {\bf  -8.15} \\
           \small \it reference: ``female''     &\small \it (-16.58,-6.69) &\small \it (-16.19,-7.13) &\small \it (-13.58,-3.17) &\small \it (-14.16,-2.38) \\
 \hline
\end{tabular}

\end{center}
\caption{Estimated parametric effects for Childhood Malnutrition data. Reference categories and credibility
intervals ($1-\alpha=0.95$) obtained by MCMC are included in italics. Significant effects are set in boldface.} \label{tab:app:linear}
\end{table*}

\section{Conclusion}

The Bayesian formulation of expectile regression outlined in this paper provides both the Bayesian counterpart to frequentist expectile regression and the expectile analogue to Bayesian quantile regression. While standard semiparametric regression specifications in expectile regression can already be handled in a frequentist setting based on iteratively weighted least squares estimation, the Bayesian formulation opens up the possibility to include more complex regression specifications such as the LASSO \citep{Alhamzawi2011} or the Dirichlet process mixture priors for random effects or Bayesian regularisation priors using a conditional Gaussian prior structure as suggested for Bayesian quantile regression in \cite{Waldmann}. Moreover, Bayesian expectile regression comprises the determination of the smoothing variances $\delta_j^2$ as an integral part of the inferential procedure and provides measures of uncertainty also for complex functionals of the model parameters. However, the asymmetric normal likelihood will usually induce a model misspecification and the impact of this misspecification will have to be studied in detail.

A further integral part of this misspecification can also be found in the interval estimates constructed from the MCMC algorithm. These fail in terms of coverage for large samples and strong heteroscedasticity while the quality of the point estimates proves satisfying. That is at least in comparison to a ``classical'' LAWS estimate, for example. In consequence, the overall questions about expectile regression remain unchanged and independent from the estimation procedure. 

Two of the main questions regarding expectile regression are the crossing of expectile curves and the interpretation of single expectiles. While non-crossing estimates exist in a frequentist setting and have been proposed in different complexity by \cite{Sobotka:2010} and \cite{Schnabel:2011}, it would be at least challenging to apply them in a boosting or Bayesian setting. Regarding the interpretation of the estimates, additional arguments to the ones presented in Section~\ref{sec:bayes} are presented by \cite{SchuWalSob:2013}. However, both questions remain in the focus of research regarding expectiles.
%
%
\section*{References}

\bibliographystyle{chicago}
\bibliography{bayes_expectiles}

\end{document}